\documentclass[fleqn,10pt]{wlscirep}

\title{Highly energetic phenomena in water electrolysis}

\author[1]{A. V. Postnikov}
\author[1]{I. V. Uvarov}
\author[2]{M. V. Lokhanin}
\author[1,3,*]{V. B. Svetovoy}
\affil[1]{Yaroslavl Branch of the Institute of Physics and Technology, Russian Academy of Sciencies, Universitetskaya 21, Yaroslavl, 150007, Russia}
\affil[2]{P. G. Demidov Yaroslavl State University, Sovetskaya 14, Yaroslavl, 150000, Russia}
\affil[3]{MESA$^+$ Institute for Nanotechnology, University of
Twente, P. O. Box 217, Enschede, 7500 AE, The Netherlands}

\affil[*]{v.svetovoy@utwente.nl}

\begin{abstract}
Water electrolysis performed in microsystems with a fast change of voltage polarity produces optically invisible nanobubbles containing H$_2$ and O$_2$ gases. In this form the gases are able to the reverse reaction of water formation. Here we report extreme phenomena observed in a millimeter-sized open system. Under a frequency of driving pulses above $100\:$kHz the process is accompanied by clicking sounds repeated every $50\:$ms or so. Fast video reveals that synchronously with the click a bubble is growing between the electrodes which reaches a size of $300\:\mu$m in $50\:\mu$s. Detailed dynamics of the system is monitored by means of a vibrometer by observing a piece of silicon floating above the electrodes. The energy of a single event is estimated as $0.3\:\mu$J and a significant part of this energy is transformed into mechanical work moving the piece. The observations are explained by the combustion of hydrogen and oxygen mixture in the initial bubble with a diameter of about $40\:\mu$m. Unusual combustion mechanism supporting spontaneous ignition at room temperature is responsible for the process. The observed effect demonstrates a principal possibility to build a microscopic internal combustion engine.
\end{abstract}

\begin{document}

\flushbottom
\maketitle

\thispagestyle{empty}

\section*{Introduction}

Standard combustion process is supported by the heat produced in the course of combustion reaction \cite{Lewis1987,Law2006}. When a volume where the process proceeds becomes small the reaction stops because of fast heat escape via the volume boundaries \cite{Veser2001,Fernandez2002}. For this reason a minimal size of microcombustors cannot be much smaller than $1\:$mm \cite{Maruta2011,Chou2011} and their volume is at least a few cm$^3$. In spite of these facts spontaneous combustion of hydrogen and oxygen was observed in microsystems in nano \cite{Svetovoy2011} and microbubbles \cite{Postnikov2016} (see also a recent review \cite{Svetovoy2016}). Mechanism of the combustion in such small volumes is not clear, but observations suggest that the bubble surface plays a role similar to a catalyst.

Nanobubbles (NBs) containing stoichiometric mixture of H$_2$ and O$_2$ gases were produced in microsystems using voltage pulses of alternating polarity (AP) \cite{Svetovoy2016}. Under sufficiently large amplitude $U > 4-5\:$V and pulses repetition frequency $f\sim 100\:$kHz a local concentration of both gases above the same electrode reaches very high values. It can be so large that the bubbles containing mixture of gases are nucleated homogeneously. Due to the homogeneous formation a large number of small bubbles emerges instead of a small number of large bubbles as it happens in the normal electrolysis \cite{Svetovoy2013}. These small bubbles do not grow larger than $200\:$nm in size because the reaction between gases is initiated spontaneously \cite{Svetovoy2011}. Although the combustion was not observed directly due to a short lifetime and a small size of the objects, a series of signatures strongly suggest that the reaction occurs.

Not all the gas produced electrochemically is burned in NBs. Some bubbles contain only oxygen or only hydrogen. These NBs survive resulting in a gradual pressure increase in a closed microchamber \cite{Svetovoy2014}. When the pulses are switched off the pressure relaxes in $100\:\mu$s i.e. much faster than the time needed to dissolve the gases in the liquid. Fast relaxation of the pressure was explained by merging of H$_2$ and O$_2$ nanobubbles with formation of a bubble containing a stoichiometric mixture of gases. The latter rapidly disappears due to the combustion reaction.

A new phenomenon emerges when concentration of the H$_2$ and O$_2$ nanobubbles becomes so large that the bubbles touch each other and merge during the process \cite{Postnikov2016}. In this case microbubles (MBs) with a typical size of $10\:\mu$m appear in the chamber as can be seen in stroboscope snapshots. Dynamics of the bubbles was too fast to observe it optically. Appearance of a MB is accompanied by a significant pressure surge in the closed chamber which lasts for a few microseconds as was measured by a vibrometer. The effect was explained by merging of many NBs with the subsequent combustion of gases in the formed MB. The energy produced in one event was estimated as $3\:$nJ.

Observation of the combustion in nano and microbubbles shows that it is possible in principle to overcome the fundamental limit for scaling down the internal combustion engines \cite{Svetovoy2016}, which can be used to power different kind of microdevices \cite{Abhari2012,Ashraf2011,Weiss2011,Volder2010}. However, the energy produced by the combustion in NBs turns mostly into heat \cite{Svetovoy2014}. Only 5\% of the combustion energy in MBs is transformed into mechanical work done by a flexible membrane covering the microchamber \cite{Postnikov2016}.

In this paper we report formation of exploding bubbles and their dynamics in a millimeter-sized system. The combustion reaction between hydrogen and oxygen in these bubbles is ignited spontaneously at room temperature.  Formation of the bubbles is accompanied by audible sounds, the energy released in the explosion is two orders of magnitude larger than was observed previously, and a significant part of this energy is transformed into useful mechanical work. On the other hand, the reaction is not initiated in the bubbles generated from an external source of the gas mixture.

\section*{Results}

\begin{figure}[ht]
\centering
\includegraphics[width=0.9\linewidth]{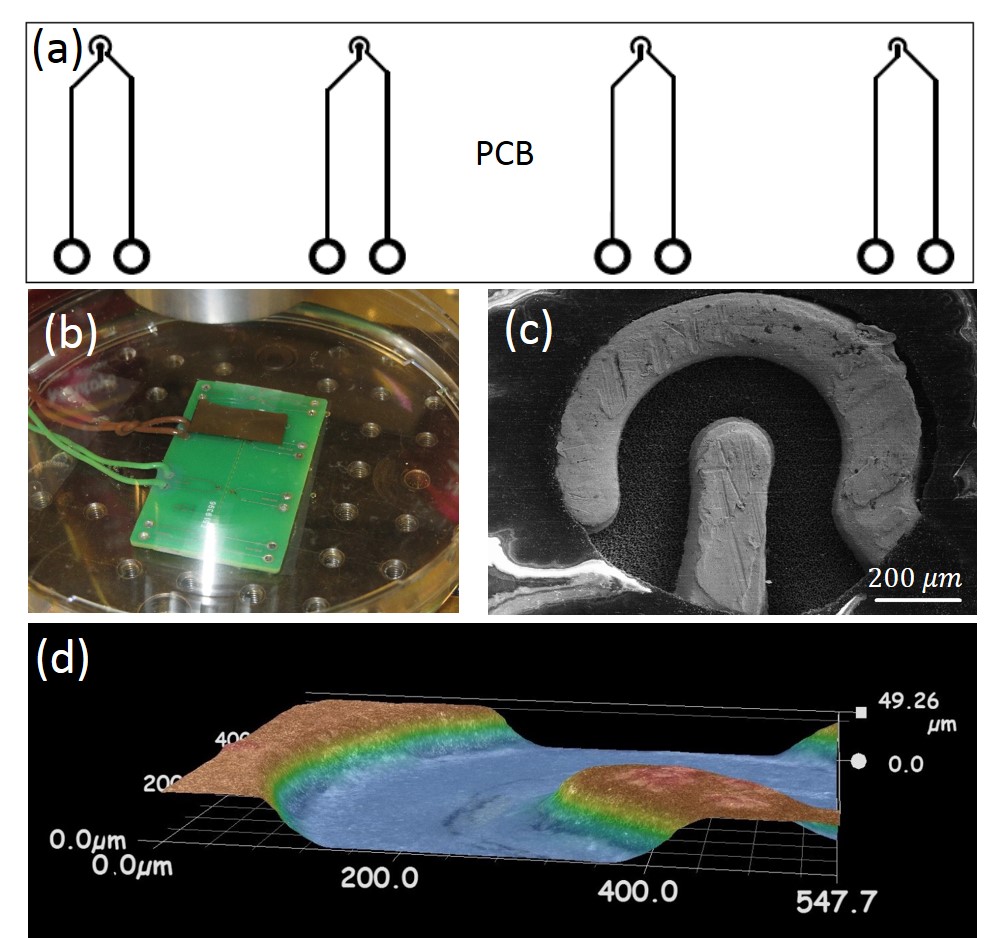}
\vspace{-0.3cm}
\caption{(a) Schematic representation of the PCB. (b) Ready-to-use device in a Petri dish. A piece of silicon is floating above one of the structures. (c) Scanning electron microscope image of the electrodes. (d) Profilometer image of the electrodes. Colors from blue to red correspond to increase of the height. \label{fig:setup}}
\vspace{-0.5cm}
\end{figure}

Figure \ref{fig:setup} shows a device used to generate gas in the system. A circular shape of the electrodes helps to better localize the produced gas. When a DC potential is applied to the electrodes we observe intense formation of well visible bubbles on both electrodes. The gases are hydrogen and oxygen since a small concentration of dissolved Cu$^{2+}$ ions does not play a significant role.  When the AP pulses are applied at a frequency of $100\:$kHz or more no gas is visible. However, the Faraday component of the current increases in comparison with that in the DC regime. This component can be separated by fitting each pulse with the function of time $I(t)=I_F+I_1e^{-t/\tau}$ (see \cite{Svetovoy2013} for detail), where $I_F$ is the Faraday current and the second term is responsible for the charging-discharging effects. For driving  pulses with the amplitude $U=6.75\:$V and the frequency $f=200\:$kHz we have found $I_F\approx 120\:$mA and $\tau\approx 1\:\mu$s.

\begin{figure}[ht]
\centering
\includegraphics[width=\linewidth]{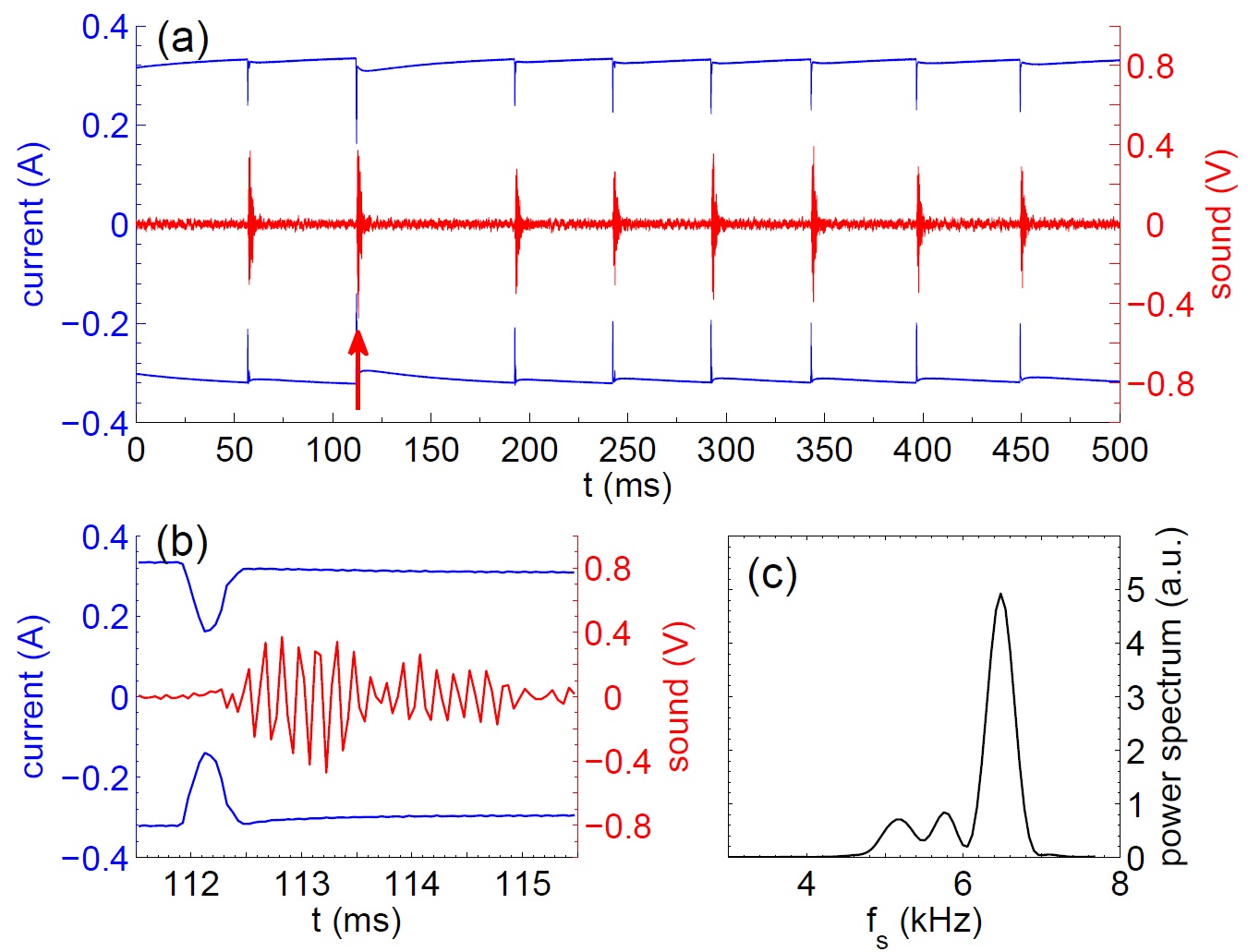}
\vspace{-0.3cm}
\caption{Sound and current at $U=6.75\:$V, $f=200\:$kHz. (a) Sound (red, right axis) from eight clicks and total current (blue, left axis) as recorded synchronously by the PicoScope. Only enveloping lines for the current are shown. A region around the red arrow is zoomed in (b), where the sound signal is averaged over $50\:\mu$s. Panel (c) shows the power spectrum as a function of sound frequency $f_s$.  \label{fig:current_sound}}
\vspace{-0.5cm}
\end{figure}

\subsection*{Sound and current}

Although no visible bubbles are produced in the AP regime, the process is accompanied by clicking sounds, which are repeated  every $50-100\:$ms.
The sound generated by the process and the total current in the system
are shown in Fig.$\:$\ref{fig:current_sound}. The amplitude of the driving voltage and its frequency were $U=6.75\:$V and $f=200\:$kHz, respectively. Only enveloping lines for the current are shown since fast oscillations cannot be resolved on the timescale of the figure. As one can see in panels (a) and (b) each click is related to a current drop lasting for about $500\:\mu$s. An expected delay of $0.4\:$ms is observed between the beginning of the current drop and the starting moment of the sound because the microphone was positioned at a distance of $12-15\:$cm from the electrodes. The clicks are nearly but not exactly periodic. Their amplitude and the interval between the clicks correlate with the depth of the current drop: the deeper the current drop is the higher is the amplitude and the longer is the interval to the next click. Frequency spectrum of the sound is shown in panel (c). It varies with the dish size and shape, depends on the proximity of the objective and other geometrical characteristics of the setup. As demonstrated in Supplementary Information (Fig.$\:$1S) separate lines of the click are essentially defined by the eigenmodes of the system. After the drop the current slowly returns to the  value it had before the drop. It is because the liquid near the structure is heated by a few degrees by the current but the bubble formation and termination bring colder liquid to the structure (see Fig.$\:$2S and explanations in Supplementary Information).

\subsection*{Fast video}

\begin{figure}[ht]
\centering
\includegraphics[width=\linewidth]{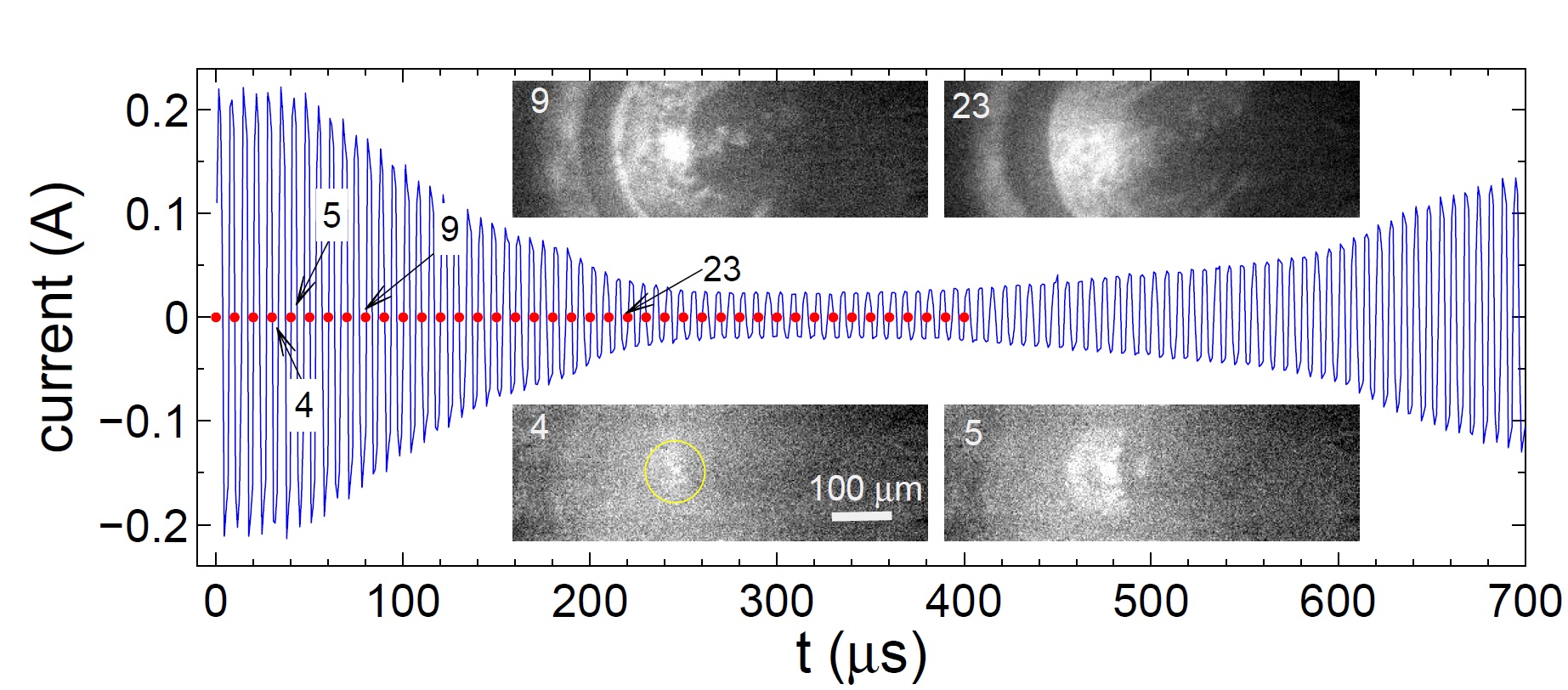}
\vspace{-0.3cm}
\caption{The current drop at $U=6.75\:$V, $f=150\:$kHz and the synchronized position of the frames (dots) in the fast video. A few frames indicated by the arrows are shown as insets. The central electrode and the emerging bubble are in the field of view. In frame 4 the initial bubble is circled.    \label{fig:current_frame}}
\vspace{-0.5cm}
\end{figure}

Fast video (see Supplementary information, file V1) reveals the reason for the current drop. A bubble growing at the periphery of the central electrode blocks the current. Figure \ref{fig:current_frame} shows the current as a function of time. Red dots indicate position of the frames in the video and a few frames are shown as insets. The initial size of the bubble can be roughly estimated from the frame 4 where the bubble appears for the first time (surrounded by a circle of diameter $100\:\mu$m). The bubble diameter is in the range of $30-50\:\mu$m. Large uncertainty is due to difficulty to define the bubble edge in this frame. Low quality of optical images is mostly due to poor reflectivity of the structure. Note that when the initial bubble appeared the current has not yet changed. The initial inflation rate is estimated from frames 4 and 5 as $6\:$m/s but it slows down with time. The bubble reaches a size of $300\:\mu$m in $50\:\mu$s (frame 9). At this point the inflation becomes slow and after frame 23 the bubble gets out of focus. It is not possible to see on this video disappearance of the bubble but for smaller bubbles, which stay in focus, one can see that the bubble shrinks and disappears (see Supplementary Information, file V2).

\begin{figure}[ht]
\centering
\includegraphics[width=\linewidth]{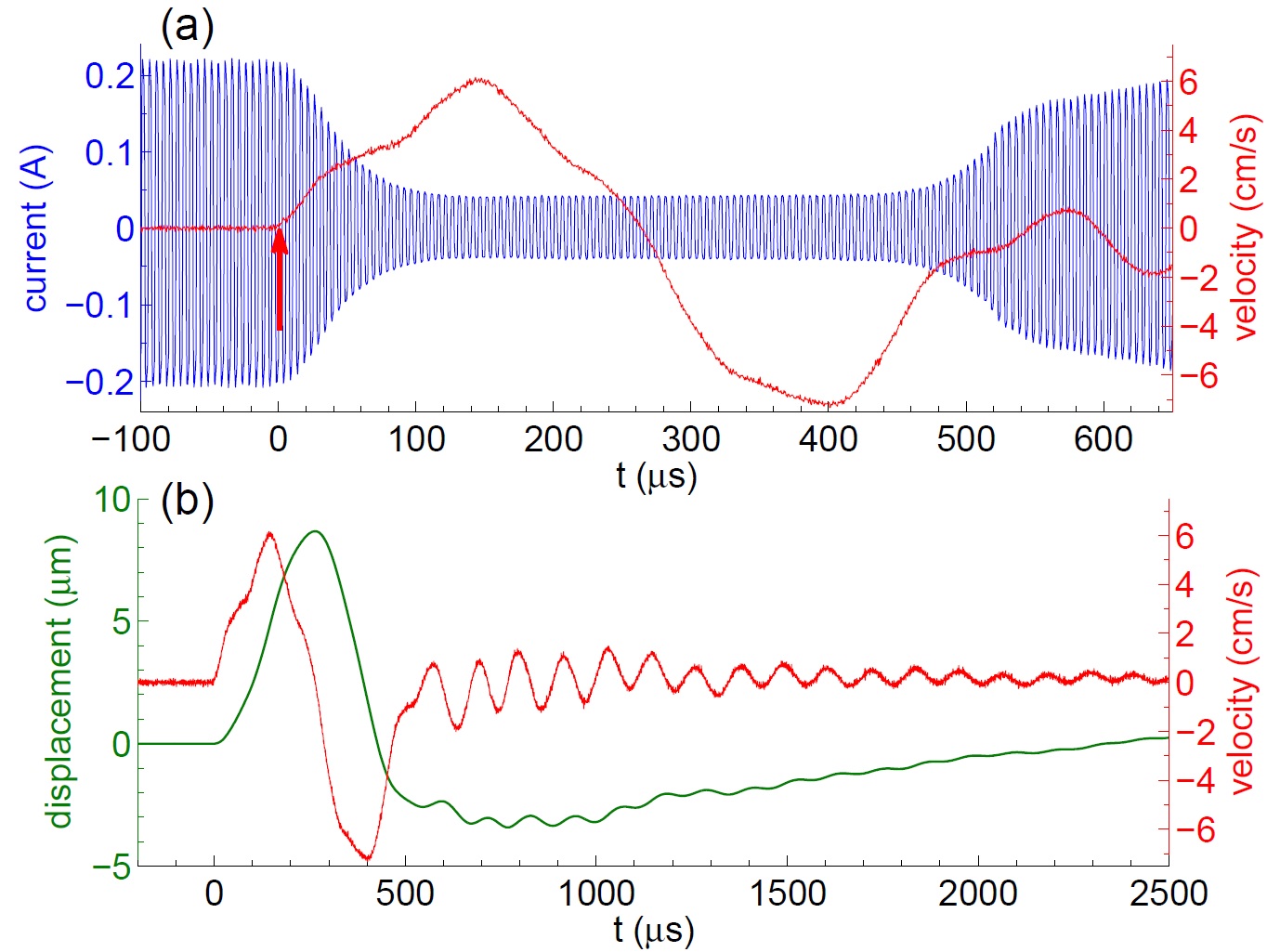}
\vspace{-0.3cm}
\caption{Movement of the floating Si-sample at $U=6.75\:$V, $f=200\:$kHz. (a) Current (left axis) and the vibrometer signal (right axis) recorded synchronously as functions of time. The arrow indicates the point where the signal is three times larger than the noise level. This point is taken as $t=0$. (b) The same vibrometer signal (velocity) on a larger time scale (right axis). The green curve (left axis) shows the displacement of the silicon piece in its center, which is just the signal integrated over time.   \label{fig:current_velocity}}
\vspace{-0.5cm}
\end{figure}

\subsection*{Vibrometer measurement}

The sound produced by the process is a sign that highly-energetic events are happening in the system. To characterize these events we use the vibrometer. A piece of silicon with dimensions of $17\times 8\times 0.5\:$mm$^3$ floats in the Petri dish with its center just above the electrodes. The laser beam is positioned on the center of the piece. The vibrometer registers velocity of the Si sample in its center. The result is presented in Fig.$\:$\ref{fig:current_velocity}(a) synchronously with the current in the system. The velocity signal is in agreement with the fast camera observations. Growing bubble moves the plate up but shrinking bubble pulls it down. However, what is striking is the magnitude of the effect.  A non-zero signal appears about $10\:\mu$s before the current starts to decrease. The initial acceleration is estimated as $710\:$m$^2$/s. Velocity of the sample reaches a maximal value of $v\approx 6\:$cm/s at the moment $t=146\:\mu$s. The signal is zero again in the middle of the current drop, where inflation of the bubble changes to deflation. The latter corresponds to negative values of the Si-sample velocity. When the bubble disappears the signal changes its behavior and oscillates with a frequency of about $10\:kHz$ as one can see in panel (b).
The oscillations are related to the lowest flexural mode of the Si-sample. For a free sample its frequency is estimated as $15\:$kHz but the floating sample is not actually free and its frequency is reduced. Since the main wavelength $\lambda=52\:$mm of the sound is much larger than the liquid thickness $H$ between the electrodes and floating silicon (estimated as $H=2-0.5=1.5\:$mm), the liquid moves together with the sample and the frequency reduction factor can be expressed via the added mass as $\left[\rho_s h/(\rho_l H+\rho_s h)\right]^{1/2}\approx 0.66$, where $h=0.5\:$mm is the thickness of the piece, $\rho_l=1\:$g/cm$^3$ and $\rho_s=2.33\:$g/cm$^3$ are the densities of the liquid and solid, respectively.Thus, the flexural frequency of the Si sample is estimated as $9.9\:$kHz.

Figure \ref{fig:current_velocity}(b) shows the velocity signal and displacement of the Si-sample on a larger timescale. The bulky piece of silicon is moved up to $9\:\mu$m by the process. The velocity of the sample measured by the vibrometer gives all the mechanical information on its movement. Using the work-energy principle we can estimate the work $W$ done by the inflating bubble on the piece as its maximal kinetic energy $W=mv^2_{max}/2\approx 0.28\:\mu$J, where $m=156\:$mg is the sample mass. This work is a lower limit for the energy of the event $E_{ev}$ because some energy dissipates and some escapes via the liquid due to the longitudinal movement. Nevertheless, we can take $E_{ev}\approx 0.3\:\mu$J as a good estimate since the quality factor is not small as oscillations in Fig.$\:$\ref{fig:current_velocity}(b) demonstrate and only a small part of the energy escapes in the longitudinal direction because the sample size is much larger than the liquid layer underneath it.

\subsection*{Microfluidic generation}
\begin{figure}[ht]
\centering
\includegraphics[width=\linewidth]{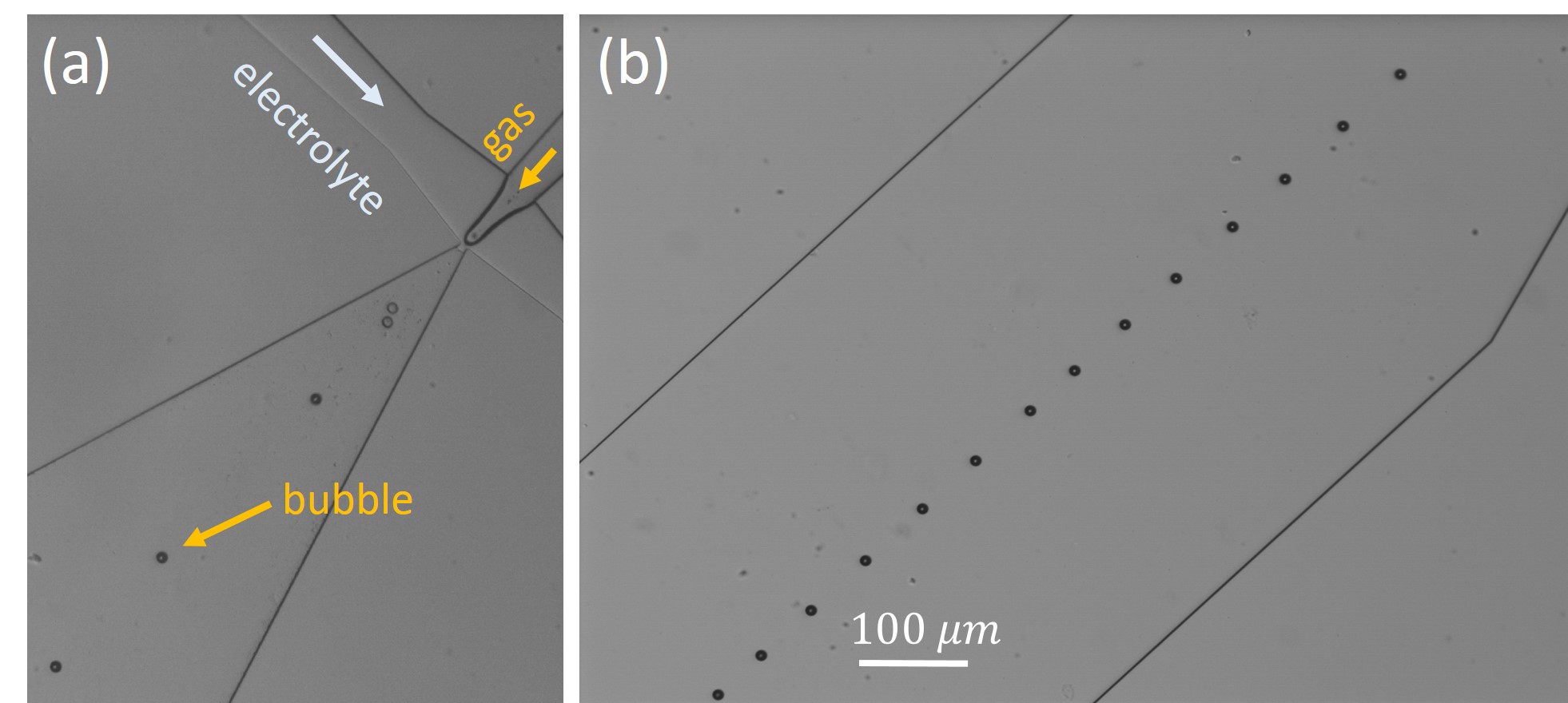}
\vspace{-0.3cm}
\caption{Generation of MBs from an external source of gas mixture. (a) Supplying channels for gas and liquid and the diffuser, where the bubbles are generated from breaking the meniscus. (b) Train of MBs containing H$_2$ and O$_2$ mixture in the electrolyte produced by the bubble generator.  \label{fig:b_generator}}
\vspace{-0.5cm}
\end{figure}

We did a special investigation to compare MBs produced by the electrochemical process and MBs generated from the external source of stoichiometric mixture of gases. In the latter case the bubbles were produced by a microfluidic bubble generator \cite{Hettiarachchi2007}. The stoichiometric gas mixture was fed into one channel of the generator while the electrolyte or clean water was fed into the other channel. The bubbles produced by the generator have a size of $10-12\:\mu$m (see Fig.$\:$\ref{fig:b_generator}). These bubbles survive at least $900\:$ms before going out of the field of view.

\section*{Discussion}

As follows from Fig.$\:$\ref{fig:current_velocity}(a) the velocity of the Si-sample builds up in less than $10\:\mu$s. Since the initial acceleration is large $\sim70g$, where $g$ is the free-fall acceleration, we conclude that the processes have an explosive character. Presence of the sound is an additional characteristic feature of the explosion. These signatures can be produced by the combustion reaction between H$_2$ and O$_2$ gases in the initial bubble. If the bubble is filled with a stoichiometric mixture of gases,  the combustion energy is estimated as $E_{com}=2N|\Delta H|/3$, where $N$ is the number of gas molecules in the bubble and $\Delta H\approx -242\:$kJ/mol is the enthalpy of water formation. The number $N$ can be expressed using the equation of state at room temperature before the combustion. Thus, we find that $E_{com}\approx 0.3\:\mu$J for the initial radius of the bubble $R_{in}\approx 22\:\mu$m. This size is in agreement with the rough estimate from the video. The ratio of the useful work to the explosion energy $W/E_{com}\sim 1$ can be compared to a similar value $W/E_{com}\simeq 0.05$ achieved in \cite{Postnikov2016}, where the work was done by the flexible membrane closing the microchamber.

Underwater combustion and explosions were investigated in many papers (see, for example, \cite{Kumakura1992,Kumakura1996,Wu2005,Klaseboer2005,Teslenko2014}) but it is difficult to compare the results with our case due to much larger scale of the events and different geometrical configuration of the experiments and modelling. Combustion in bubbles as small as $10\:\mu$m in diameter was considered in \cite{Nguyen2005} but it was supported by the high temperature inside the bubble rapidly collapsing in an acoustic field.

Direct comparison can be done with the experiment \cite{Teslenko2010} where combustion was observed in a bubble (in water) of $2\:$mm in diameter containing acetylene-oxygen mixture. A significant energy of $20\:$mJ was supplied by a spark to ignite the normal combustion in the bubble. In our case the reaction is initiated spontaneously in a smaller bubble but the rates of inflation are comparable: $6\:$m/s in our case vs $8\:$m/s in \cite{Teslenko2010}. An additional relevant quantity is the energy density. In our case it can be estimated directly from the experiment as $E_{ev}/V_{in}=5-21\:$MJ/m$^3$, where $V_{in}=4\pi R_{in}^3/3$ is the initial volume of the bubble. Wide interval here is due to uncertainty of the initial bubble size. If we use $E_{com}$ instead of the experimental value $E_{ev}$ the volume $V_{in}$ falls out and we find the theoretical value $6.5\:$MJ/m$^3$ for the hydrogen-oxygen combustion. This energy density can be compared with $15\:$MJ/m$^3$ for the acetylene-oxygen mixture. It differs from that for hydrogen-oxygen because the enthalpy of the acetylene combustion is rather large, $\Delta H_{C_2H_2}=-1318\:$kJ/mol. From the comparison one can conclude that we really observe combustion of hydrogen in microbubbles.

The initial bubble containing stoichiometric mixture of gases is formed by merging of the NBs produced by the alternating polarity electrochemical process \cite{Postnikov2016}. For this, the density of NBs must be so high that the bubbles are nearly touching. This is why the events are separated by a time of about $50\:$ms as can be seen in Fig.$\:$\ref{fig:current_sound}. After explosion the system waits while new NBs will be collected near the structure. Fluctuations of the local concentration of nanobubbles explain that the clicks occurred not completely regularly.

A reason for the spontaneous initiation of the reaction in a relatively large microbubble is not clear. In NBs containing mixture of gases the reaction could be initiated as a surface-assisted process \cite{Svetovoy2016}, but for larger bubbles this mechanism has to be suppressed due to a smaller surface-to-volume ratio. The stoichiometric bubbles produced by the microfluidic generator survive at least $900\:$ms. On the other hand, in the bubbles formed by coalescence of NBs the reaction is initiated spontaneously in less than $10\:\mu$s. We do not know the reason for this difference but it could be related to nanodroplets left in the bubble after merging of many NBs. Densely packed NBs fill only 74\% of volume, the rest is the liquid trapped in between the bubbles. After coalescence of the bubbles the trapped liquid will form nanodroplets distributed inside of the MB. In this MB any gas molecule is separated from the liquid by a nanoscopic distance. Possibly it can play a role for the initiation of the reaction.

When the explosion happens in the initial MB, the pressure and temperature in the bubble rise.  If  heat exchange with the walls is slow in comparison with the reaction, the combustion energy $E_{com}$ is spent only on heating of the water vapor being formed in the reaction.  Assuming that the bubble size changes insignificantly during the combustion phase we can find the pressure in the bubble from the equation of state as $P/P_0 \approx 2|\Delta H|/3T_0=64.7$, where $P_0\approx 1\:$bar is the ambient pressure and $T_0$ is the room temperature. The most efficient channel for the heat exchange is water vaporization from the walls by energetic molecules. For such molecules a time needed to reach the wall is estimated as $\tau_h\sim R_{in}^2/D_{gg}$, where $D_{gg}$ is the self-diffusion coefficient for water vapor. Under normal conditions $D_{gg}\approx 2.8\times 10^{-5}\:$ m$^2$/s and we find $\tau_h \sim 10\:\mu$s; this time is scaled with temperature as $(T/T_0 )^{1/2}$. Spontaneous combustion in micro and nanobubbles happens in a few microseconds \cite{Svetovoy2016}. Since both the time for the heat exchange and the time for combustion are comparable the pressure and temperature in the initial bubble immediately after the combustion is defined by the dynamics but anyway it is expected to be high. The source of the sound is the pressure surge.
During the inflation phase the pressure decreases and going below the atmospheric pressure by inertia as it happens, for example, in cavitation. At the final stage the bubble shrinks and disappears. The time scale for heat exchange explains why the useful work observed in this study is much larger than that in the microsystem \cite{Postnikov2016}. In the latter case the initial bubble had a radius of $R_{in}\approx 5\:\mu$m. Due to smaller size the heat exchange happens much faster and the pressure and temperature reach smaller values.

Alternatively, one could try to explain the observed phenomena by heating of the electrolyte by the Faraday current (Joule heating). A vapor bubble could be formed due to the heating. However, the heating of the electrolyte has to result in the current increase. This effect can be used as a build-in thermometer \cite{Svetovoy2014}. What is observed is only a small temperature increase between the clicks as explained in Supplementary Information, Fig. 2S. In this scenario there is no a driving force for the bubble inflation since pressure in the bubble cannot reach a high value. As an additional scenario one could imagine that the current passing through a small liquid volume vaporizes this volume producing the initial bubble with a high pressure inside. The number of molecules in the bubble is estimated as $N_v < E_{ev}/\Delta H_v\approx 4.4\times 10^{12}$, where $E_{ev}\approx 0.3\:\mu$J is the observed energy and $\Delta H_v=41\:$kJ/mol is the enthalpy of water vaporization. We took only the upper limit because the heating of the molecules was neglected. For this $N_v$ the radius of vaporized liquid sphere must be $R_l<3.2\:\mu$m. The experimental value of the Faraday current density is estimated as $j_F\approx 200\:$A/cm$^2$. It is similar to that used for normally
functioning microdevices \cite{Svetovoy2013,Svetovoy2014}. This current can transfer the energy $E_{ev}\approx 0.3\:\mu$J to $N_v$ molecules for the time $\tau_E$, which can be found from the relation $\pi R_l^2 j_F U\tau_E = E_{ev}$, where $U=6.75\:$V. It gives $\tau_E > 700\:\mu$s that is considerably larger than the time needed to produce the initial bubble, which is smaller than $10\:\mu$s.

For application in microcombustors it is not efficient to produce exploding gas electrochemically. Instead,  one has to provide appropriate conditions in a microvolume with premixed gases. If the volume is closed by a flexible membrane, the explosion will move the membrane up. Therefore, a critical step is to understand why the gas is ignited spontaneously in MBs produced electrochemically and ensure the right conditions in the microvolume.

In conclusion, we observed highly energetic events in water electrolysis produced by short voltage pulses of alternating polarity. The process is accompanied by the sound clicks, which occur synchronously with the current drops in the system. The observations were explained by spontaneous combustion of H$_2$ and O$_2$ gases in the initial bubble with a diameter of about $40\:\mu$m. The combustion produces the pressure jump, which generates sound and drives the bubble inflation. Unusual reaction mechanism has to drive the process to provide spontaneous ignition at room temperature. Significant part of the combustion energy was transformed into mechanical work instead of heat. It opens a practical way to build a truly microscopic internal combustion engine.

\section*{Methods}

Standard printed circuit boards (PCB) with Cu foil is used for bubble generation in the AP regime. One PCB contains four pairs of electrodes and contact lines as shown in Fig.$\:$\ref{fig:setup}. Typical size of the circular structure is $1\:$mm, the line width is $150\:\mu$m, and the line thickness is $50\:\mu$m. Only the circular structures are in the electrical contact with the electrolyte, the rest is covered with a standard insulating coating used for PCB.

The PCB device is placed in a Petri dish and covered by 2-3 mm of the electrolyte ($1\:$mol/l solution of Na$_2$SO$_4$ in water). One electrode is kept grounded but the other one is at the alternating potential with an amplitude of $6-7\:$V. A relatively high potential is used to produce a significant supersaturation with H$_2$ and O$_2$ gases. The voltage pulses were generated by a PicoScope 5000 and amplified 15 times. The current in the system and the produced sounds were recorded using different channels of the PicoScope. On a long time scale ($\geq 1\:$s) Windows Sound Recorder was used. The sound was recorded in air with a microphone installed $10-12\:$cm from the Petri dish. One of the sound files is provided in Supplementary Information. Visual dynamics of the system was analysed with a fast camera Photron Fastcam SA1.1 at the frame rate up to $100\:000\:$fps. A detailed analysis of the process was performed with a vibrometer Polytec MSA-400.



\section*{Acknowledgements}

We thanks D. Lohse, A. Prosperetti, S. Wildeman for helpful discussions and G. Lajoinie, M. van Limbeek, R. Sanders for technical assistance. This work is supported by the Russian Science Foundation, grant 15-19-20003. V.B.S. acknowledges support from the Netherlands Center for Multiscale Catalytic Energy Conversion (MCEC), an NWO Gravitation programme funded by the Ministry of Education, Culture and Science of the government of the Netherlands.

\section*{Author contributions statement}

V.B.S. conceived the experiments, A.V.P. and I.V.U. conducted the experiments, the data was analysed by M.V.L. and V.B.S. All authors reviewed the manuscript.

\section*{Additional information}
\textbf{Supplementary information} accompanies this paper at http://www.nature.com/articles/srep39381

\noindent\textbf{Competing financial interests:}
The authors declare no competing financial interests.

\end{document}